\documentclass[journal=nalefd,manuscript=letter]{achemso}

\usepackage{chemformula} % Formula subscripts using \ch{}
\usepackage[T1]{fontenc} % Use modern font encodings

\author{Xiaoran Liu}
\email{xiaoran.liu@rutgers.edu}
\affiliation[Rutgers]
{Department of Physics and Astronomy, Rutgers University, Piscataway, New Jersey 08854, USA}
\author{B. J. Kirby}
\affiliation[NIST]
{NIST Center for Neutron Research, National Institute of Standards and Technology, Gaithersburg, Maryland 20899, USA}
\author{Zhicheng Zhong}
\affiliation[Ningbo]
{Ningbo Institute of Materials Technology and Engineering, Chinese Academy of Sciences, Ningbo, Zhejiang 315201, China}
\author{Yanwei Cao}
\affiliation[Ningbo]
{Ningbo Institute of Materials Technology and Engineering, Chinese Academy of Sciences, Ningbo, Zhejiang 315201, China}
\author{B. Pal}
\affiliation[Rutgers]
{Department of Physics and Astronomy, Rutgers University, Piscataway, New Jersey 08854, USA}
\author{M. Kareev}
\affiliation[Rutgers]
{Department of Physics and Astronomy, Rutgers University, Piscataway, New Jersey 08854, USA}
\author{S. Middey}
\affiliation[IIOS]
{Department of Physics, Indian Institute of Science, Bengaluru 560012, India}
\author{J. W. Freeland}
\affiliation[APS]
{Advanced Photon Source, Argonne National Laboratory, Argonne, Illinois 60439, USA}
\author{P. Shafer}
\affiliation[ALS]
{Advanced Light Source, Lawrence Berkley National Laboratory, Berkeley, California 94720, USA}
\author{E. Arenholz}
\affiliation[ALS]
{Advanced Light Source, Lawrence Berkley National Laboratory, Berkeley, California 94720, USA}
\author{J. Chakhalian}
\affiliation[Rutgers]
{Department of Physics and Astronomy, Rutgers University, Piscataway, New Jersey 08854, USA}

\title[An \textsf{achemso} demo]
  {Emergent magnetic state in (111)-oriented quasi-two-dimensional spinel oxides}

\abbreviations{IR,NMR,UV}
\keywords{spinels, ultrathin films, emergent properties, magnetism}

\begin{document}

%%%%%%%%%%%%%%%%%%%%%%%%%%%%%%%%%%%%%%%%%%%%%%%%%%%%%%%%%%%%%%%%%%%%%
%% The "tocentry" environment can be used to create an entry for the
%% graphical table of contents. It is given here as some journals
%% require that it is printed as part of the abstract page. It will
%% be automatically moved as appropriate.
%%%%%%%%%%%%%%%%%%%%%%%%%%%%%%%%%%%%%%%%%%%%%%%%%%%%%%%%%%%%%%%%%%%%%
%\begin{tocentry}

%Some journals require a graphical entry for the Table of Contents.
%This should be laid out ``print ready'' so that the sizing of the
%text is correct.

%Inside the \texttt{tocentry} environment, the font used is Helvetica
%8\,pt, as required by \emph{Journal of the American Chemical
%Society}.

%The surrounding frame is 9\,cm by 3.5\,cm, which is the maximum
%permitted for  \emph{Journal of the American Chemical Society}
%graphical table of content entries. The box will not resize if the
%content is too big: instead it will overflow the edge of the box.

%This box and the associated title will always be printed on a
%separate page at the end of the document.

%\end{tocentry}

%%%%%%%%%%%%%%%%%%%%%%%%%%%%%%%%%%%%%%%%%%%%%%%%%%%%%%%%%%%%%%%%%%%%%
%% The abstract environment will automatically gobble the contents
%% if an abstract is not used by the target journal.
%%%%%%%%%%%%%%%%%%%%%%%%%%%%%%%%%%%%%%%%%%%%%%%%%%%%%%%%%%%%%%%%%%%%%
\begin{abstract}
We report on the emergent magnetic state of (111)-oriented CoCr$_2$O$_4$ ultrathin films sandwiched by Al$_2$O$_3$ in the quantum confined geometry. At the two-dimensional crossover, polarized neutron reflectometry reveals an anomalous enhancement of the total magnetization compared to the bulk value. Synchrotron x-ray magnetic circular dichroism (XMCD) demonstrates the appearance of long-range ferromagnetic ordering of spins on both Co and Cr sublattices. Brillouin function analyses further corroborates that the observed phenomena are due to the strongly altered magnetic frustration, manifested by the onset of a Yafet-Kittel type ordering as the new ground state in the ultrathin limit, which is unattainable in the bulk.  
\end{abstract}

%%%%%%%%%%%%%%%%%%%%%%%%%%%%%%%%%%%%%%%%%%%%%%%%%%%%%%%%%%%%%%%%%%%%%
%% Start the main part of the manuscript here.
%%%%%%%%%%%%%%%%%%%%%%%%%%%%%%%%%%%%%%%%%%%%%%%%%%%%%%%%%%%%%%%%%%%%%

%\section{Introduction}

The quest to design, discover and manipulate new quantum states of matter has fostered tremendous research activity among condensed matter physicists. Recent progress in the fabrication of epitaxial thin films has empowered this effort with additional means and led to a plethora of interesting artificial multilayers and heterostructures grown with atomic level of precision \cite{Schlom_ARMR_2007,Hwang_Nature_2012,Stemmer_ARMR_2014,Anand_ARMR_2014}. Nowadays, to realize exotic physics linked to many-body  phenomena the interest has shifted to tailoring the magnetic states in quasi two-dimensional (2D) limit \cite{Coey_MRS_2013,Hellman_RMP_2017}. On one hand, according to the Mermin-Wagner theorem, in an isotropic Heisenberg spin system of dimensionality $D\leq$2, enhanced thermal fluctuations prohibit the onset of a long-range (ferro- or antiferro-) magnetic ordering at any finite  temperature\cite{MW_PRL_1966}. On the other hand, lowering the dimensionality brings about several new factors that can radically alter a quantum system including changes in band topology, ionic coordinations and covalency, crystal fields, exchange pathways, magnetic anisotropy, quantum confinement, and universality class \cite{Bansil_RMP_2016,Hwang_Nature_2012,JC_RMP_2014,XL_MRS_2016,Song_NanoTec_2018}. As a result, in the crossover to low dimensions the magnetic ground state of a material can be distinctly different from its three-dimensional (3D) counterpart thus opening an opportunity for emergent or hidden materials phases. 

In this context, it is interesting to ask whether we can ``dial-in'' dimensionality of a system from 3D to 2D in a controllable way, and what can happen to the quantum state when low dimensionality entwines with frustration? Here we recap that frustrated magnets are systems where the localized spins are entangled in an incompatible way due to either multiple competing exchange interactions, or the underlaying lattice geometry or both\cite{Ramirez_review_1994,Greedan_JMC_2001,Bramwell_Science_2001,Balents_Nature_2010,Wosnitza_RPP_2016}. Generally, frustration tends to suppress spin ordering and promotes a complex magnetic phase diagram typically with a set of competing ground states \cite{Vojta_RPP_2018}. Further, a large number of theory proposals have recently addressed another aspect showing how dimensionality effectively tunes the many-body function, either driving the ground state into an entirely different regime on the magnetic phase diagram, or inducing unconventional phases through  quantum criticality \cite{Vojta_RPP_2018, Starykh_RPP_2015, Schmidt_PR_2017}.

Among many magnetically frustrated compounds, the family of chromate spinels MCr$_2$O$_4$ (M = Mn, Fe, and Co) has attracted intense interest \cite{Lee_JPSJ_2010, Tokura_AM_2010, Ganguly_PRB_2015,Windsor_PRB_2017,Heuver_PRB_2015,Aqeel_PRB_2015,Kim_APL_2009,Zhang_APL_2015,Yang_JPD_2012,Guzman_PRB_2017,Chen_JAP_2013,Pronin_PRB_2012,Kocsis_PRB_2013,Kamenskyi_PRB_2013,Efthimiopoulos_PRB_2015,Tsurkan_PRL_2013}. These materials crystallize into the normal spinel structure AB$_2$O$_4$, with M$^{2+}$ and Cr$^{3+}$ ions occupying the tetrahedral (A) and octahedral (B) sites, respectively. As both nearest-neighbor exchange interactions \textit{J}$_{\textrm{AB}}$ and \textit{J}$_{\textrm{BB}}$ are antiferromagnetic [see Fig.~1(c)], a strong competition between the exchange interactions ($J_{\textrm{BB}}/J_{\textrm{AB}} > 2/3$) causes magnetic frustration and results in a unique three-sublattice ferrimagnetic spiral order \cite{LKDM_JAP_1961,LKDM_PR_1962}. In the bulk such a conical spiral order engenders a macroscopic spontaneous polarization which is switchable by an external magnetic field \cite{Yamasaki_PRL_2006, Choi_PRL_2009, Dey_PRB_2014}, consistent with the calculations based on the spin-current model \cite{Katsura_PRL_2005} and the inverse Dzyaloshinskii-Moriya interaction model \cite{Sergienko_PRB_2006}. 

Strikingly, when viewed along the [111] direction, the spinel crystal structure shows a stacking of triangular and Kagome cation planes with intrinsically large geometrical frustration (GF), embedded in the oxygen cubic close-packed frame, as illustrated in Fig.~1(a)-(b). Specifically, in this viewpoint the basic structural arrangement is ``-O$_4$-B$_3$-O$_4$-A-B-A-'', containing four cation layers, which we denote as one quadruplet layer or QL. Therefore, we may speculate that if the lattice can be confined along such a GF direction, the magnetic frustration will be strongly altered and lead to the formation of emergent magnetic phases. Naturally, the thin film approach can give direct access to modalities for manipulation and control of these phases.

\begin{figure}[t]
\includegraphics[width=0.5\textwidth]{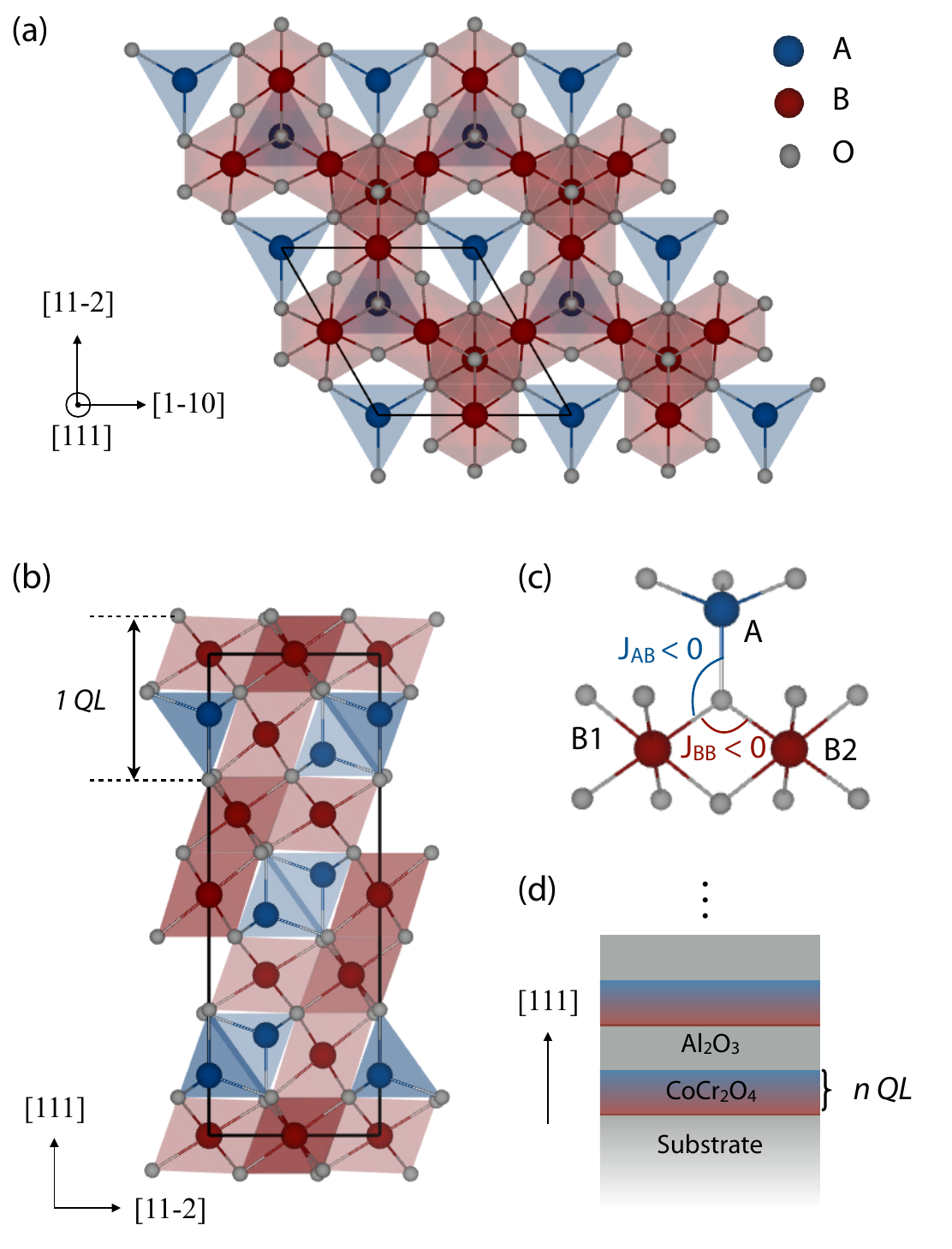}
\caption{\label{pnr} (a) Crystal structure of normal spinel AB$_2$O$_4$ viewed along [111] direction. The oxygen ions stack in the cubic close-packed framework, forming both tetrahedral and octahedral interstitials, which are separately occupied by A and B cations. (b) View of the structure along the [1-10] direction. (c) Nearesst-neighbor exchange interactions in normal spinels with magnetic A and B ions. Both J$_\textrm{AB}$ and J$_\textrm{BB}$ are antiferromagnetic, leading to magnetic frustration. (d) Schematic of the ultrathin (111)-oriented CoCr$_2$O$_4$ thin films confined by Al$_2$O$_3$ spacers.}
\end{figure}

Based on the above-mentioned framework, we demonstrate the power of this approach for ground state manipulation in the prototypical case of CoCr$_2$O$_4$ spinel layered along [111] direction. In bulk CoCr$_2$O$_4$, the ground state has the three-sublattice ferrimagnetic spiral configuration with the onset of ferrimagnetism at T$_\textrm{C}$ $\sim$ 93 K, and the incommensurate spin-spiral order at T$_\textrm{S}$ $\sim$ 26 K \cite{LKDM_PR_1962,Tomiyasu_PRB_2004}. The incommensurate to commensurate lock-in transition further takes place at T$_\textrm{L}$ $\sim$ 14 K \cite{Chang_JPCM_2009}. In this Letter, we report on the discovery of an emergent magnetic state in (111)-oriented CoCr$_2$O$_4$ ultrathin films confined by inert Al$_2$O$_3$ layers. Spin polarized neutron reflectivity (PNR) confirm the establishment of long-range magnetic ordering in CoCr$_2$O$_4$ slabs with thickness of few nanometers only. Importantly, analysis on the x-ray magnetic circular dichroism (XMCD) reveals that even though the ordering of these quasi-2D CoCr$_2$O$_4$ films is still ferrimagnetic, it is no longer of the spin-spiral type, but rather the new Yafet-Kittel type, which was theoretically proposed \cite{YK_PR_1952} but never realized in the bulk phase diagram.

%\section{Results and discussion}

(111)-oriented [$n$QL CoCr$_2$O$_4$/1.3nm Al$_2$O$_3$]$_4$ superlattices (1QL $\approx$ 0.48 nm; $n$ = 4, 2) were fabricated by pulsed laser deposition on single crystalline (0001)-oriented Al$_2$O$_3$ substrate, as sketched in Fig.~1(d). Al$_2$O$_3$ was selected as the non-magnetic confinement spacer because of the good structural compatibility with CoCr$_2$O$_4$ \cite{Xiaoran_APL_2015}. Details of the materials synthesis, structural and chemical characterizations are given elsewhere \cite{Xiaoran_APL_2014,Xiaoran_APL_2015}. Additionally, the degree of cation distribution disorder is investigated by resonant X-ray absorption spectroscopy (XAS). Within the experimental limit, no signs of cation distribution disorder or ion valency change is observed for all samples. Specifically, the obtained absorption line-shapes as well as the absorption energy peak positions at the $L_{3,2}$ absorption edges of both Co and Cr are practically identical to that of the bulk CoCr$_2$O$_4$ reference \cite{XL_MRS_2016}  (see Fig. S1, Supplementary). These results further confirm that the ultra-thin heterostructures are of expected thickness, orientation, and proper local chemical environments.

\begin{figure}[t]
\includegraphics[width=0.95\textwidth]{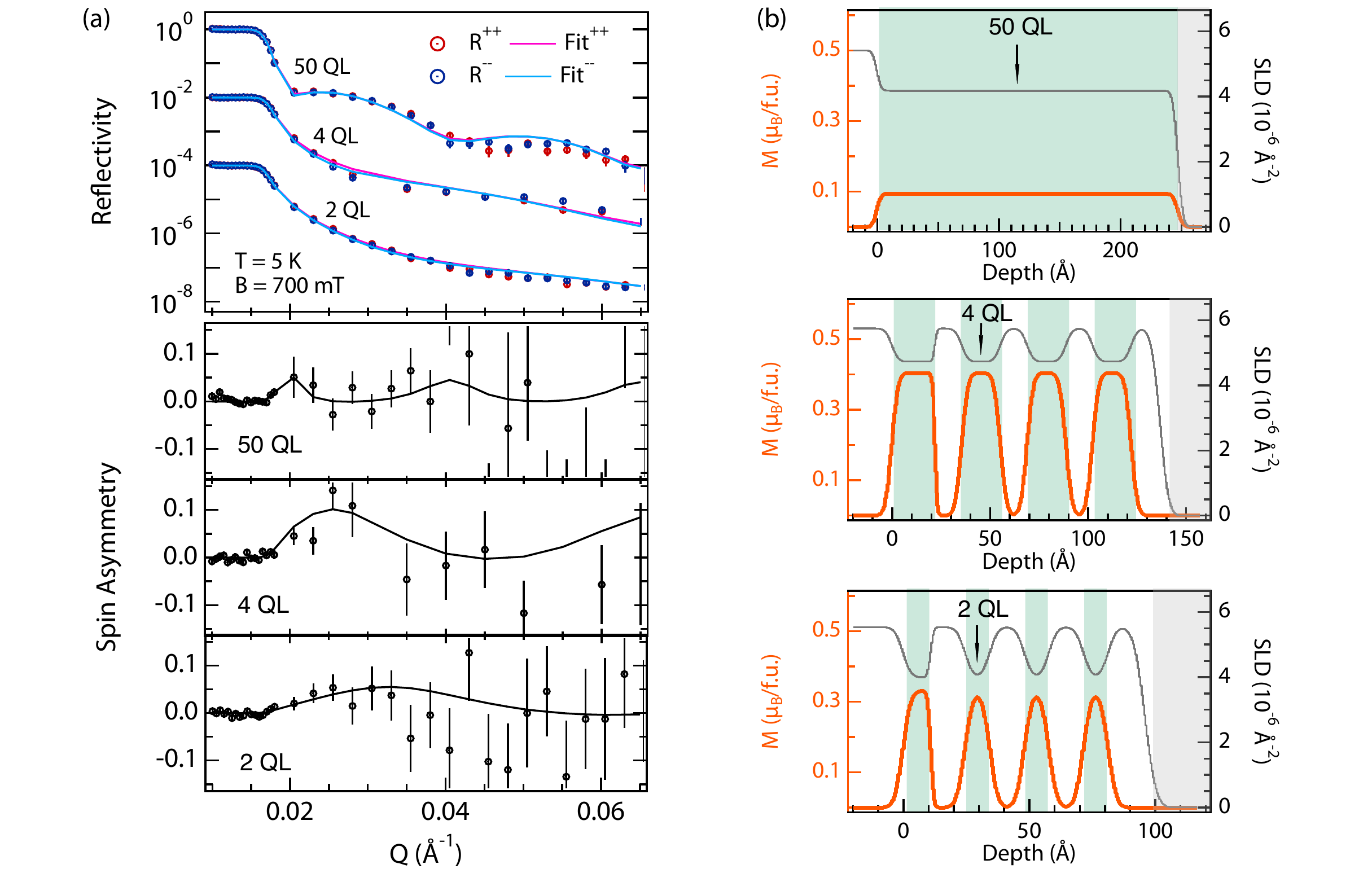}
\caption{\label{pnr} (a) Non-spin-flip PNR data with model fitting for both CoCr$_2$O$_4$ thick film (50 QL) and ultrathin [$n$QL CoCr$_2$O$_4$ / 1.3nm Al$_2$O$_3$]$_4$ superlattices, plotted as reflectivity and spin asymmetry, respectively. Reflectivity curves are offset in intensity for clarity of presentation. Spin asymmetry is defined as $(R^{++} - R^{--})/(R^{++} + R^{--})$. All data were measured at 5 K under 0.7 T in-plane magnetic field. (b) Depth profiles of the net magnetization ($M$) and the nuclear scattering length density (SLD). The regions in light green denote the CoCr$_2$O$_4$ slabs inside the sample, and the gray regions denote air on top of sample surface.}
\end{figure}

First, we discuss the presence and distribution of the net magnetization in the (111)-oriented superlattices and compare it to a bulk-like CoCr$_2$O$_4$ (50 QL, 24 nm) sample. In order to probe the rather small signals and determine the magnetic depth profiles in the superlattices, we performed the PNR experiments at the PBR beamline at the NIST Center for Neutron Scattering. The polarized neutron beam was set incident at a grazing angle and the specular reflectivity was recorded as a function of the transfer wave vector Q$_z$ along the surface normal. Depth profiles of the nuclear scattering length density (SLD) and the magnetization ($M$) component parallel to H were deduced by fitting the non spin-flip data to a superlattice model where the individual layer thickness and roughness were pre-determined from X-ray reflectivity and fixed during the fit \cite{Xiaoran_APL_2015}. Data was fitted using the NIST Refl1D \cite{refl1d} software routines. 

Figures~2(a)-2(b) present the fitted PNR data, along with the model profiles corresponding to the fits. As clearly seen, a distinct magnetization per formula unit (f.u.) is observed in each ultrathin superlattice. Note, since during the measurements we applied only a moderate magnetic field of 0.7 T, the fitted $M$ implies the presence of a spontaneous long-range magnetization rather than canting of the local moment in a paramagnetic phase. First, the validity of the model is tested on the 50 QL film, which yields $M\sim$ 0.10 $\mu_B/$f.u. very close to the reported value in bulk CoCr$_2$O$_4$ compounds with the spiral spin state \cite{Choi_PRL_2009}. Contrary to expectation, in the ultrathin case the magnitude of $M$ becomes remarkably \textit{enhanced}, reaching $\sim$ 0.39 $\mu_B$/f.u. in 4 QL and $\sim$ 0.31 $\mu_B$/f.u. in 2 QL samples, respectively. To appreciate this result, we emphasize that even in the bulk the saturated magnetization from the collinear component of the spiral order can reach only $\sim$ 0.15 $\mu_B$/f.u. \cite{Yamasaki_PRL_2006,Choi_PRL_2009}. This strongly suggests that the nearly fourfold increase of $M$ in 4 QL and 2 QL films cannot be attributed to mundane changes in magnetic anisotropy with thickness. Instead, these findings imply the presence of a more fundamental modification of the magnetic structure which takes place in the quasi 2D limit of the (111)-oriented ultrathin films.  

\begin{figure}[t]
\includegraphics[width=0.6\textwidth]{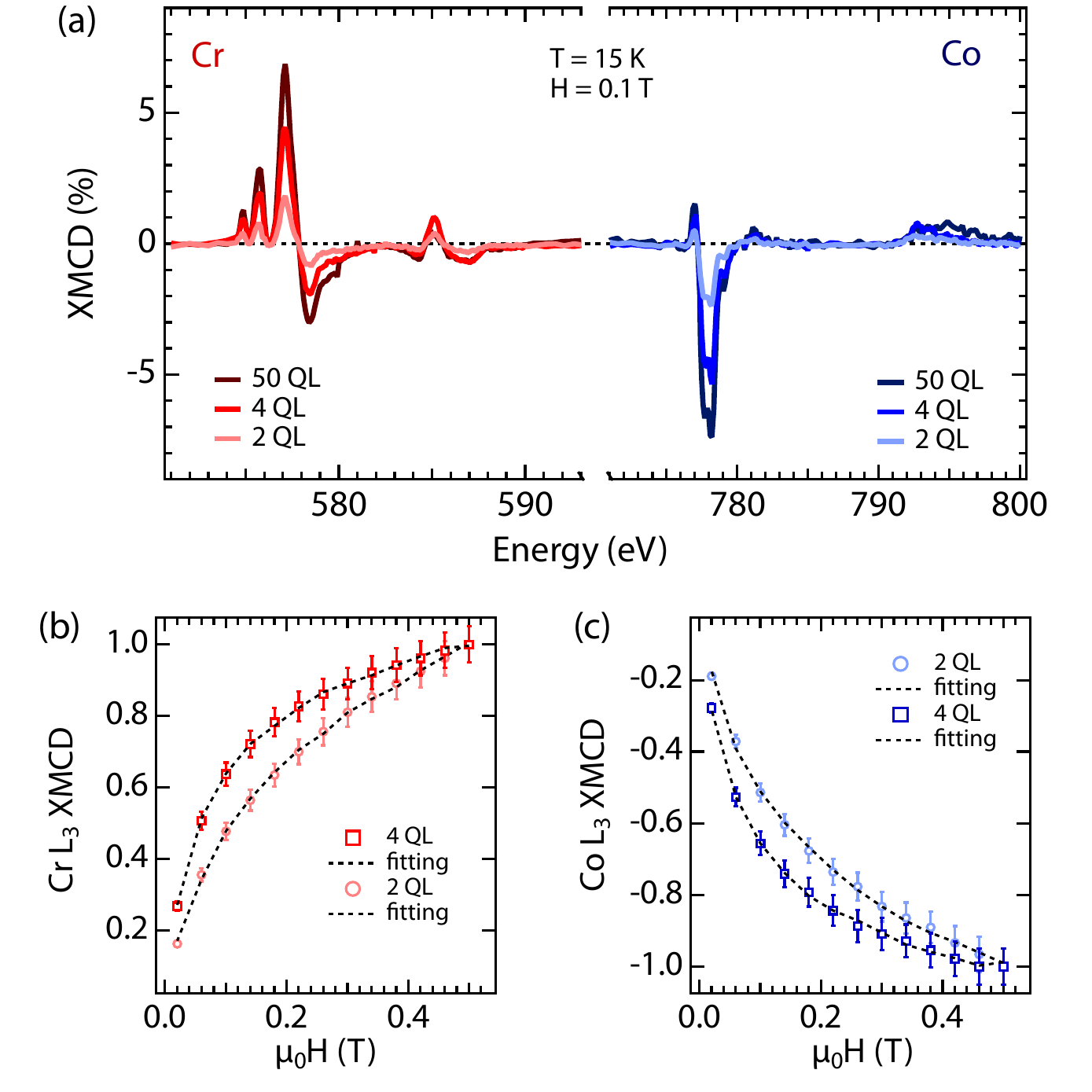}
\caption{\label{xmcd} (a) Cr and Co $L_{2,3}$ edges XMCD data of (111)-oriented CoCr$_2$O$_4$ thin films of various thickness. (b)-(c) Field-dependent XMCD results of $n$ = 4 and 2 QL samples taken at Cr and Co $L_3$ maximal peak positions (Cr at $\sim$577 eV; Co at $\sim$778 eV) with Brillouin function fittings.}
\end{figure} 
           
In order to elucidate the magnetic structure of each sublattice, we performed resonant XAS measurements with left- and right-circularly polarized beams at beamline 4.0.2 of the Advanced Light Source at Lawrence Berkeley National Laboratory. The spectra were measured at 15 K under 0.1 T magnetic field, and recorded using the luminescence detection mode. The circularly polarized x-rays were incident with an angle of 30$^\circ$ relative to the sample surface. The intensities were normalized with respect to their corresponding absorption spectra. The difference between these two spectra, known as the X-ray magnetic circular dichroism (XMCD), originates from the local magnetization of a specifically probed chemical element (i.e. Co or Cr). The XMCD results at Co and Cr L$_{3,2}$ edges are shown in Fig.~3(a). The dichroic signals of similar lineshpae are clearly evident for all samples on both elements. Moreover, the sign of the XMCD spectra is opposite for Co and Cr as observed at the strongest feature near their L$_3$ edge (Cr at $\sim$577 eV and Co at $\sim$778 eV), signifying that the spin orientation on Cr ions is antiparallel to that of the Co ions. To quantify the values of the orbital and spin magnetic moments on each element, we applied the ``sum rules'' analysis to the spectra \cite{Thole_PRL_1992,Carra_PRL_1993,Chen_PRL_1995}. The obtained results are summarized in Table~I. As seen, for all samples the magnetic moment of Co dominates over the magnetic moment of Cr and determines the overall direction of the net magnetization (i.e., $M_{\textrm{net}}$ = $M_{\textrm{Co}} + 2M_{\textrm{Cr}}$). In addition, $M_{\textrm{net}}$ exhibits strong enhancement from $\sim$0.2(3) $\mu_B$/f.u. in 50 QL to $\sim$0.3(4) $\mu_B$/f.u. in 4 QL, but reduces back to $\sim$0.1(5) $\mu_B$/f.u. in 2 QL. The non-monotonic trend of $M$ vs. $n$ is similar in both XMCD and PNR characterizations. Together these observations affirm that even in the ultrathin limit, the ground state is indeed ferrimagnetic.  

Next, we turn our attention to the spin configuration of the ferrimagnetic state in (111) CoCr$_2$O$_4$ ultrathin films. For this purpose we recorded the XMCD intensity of each element as a function of applied magnetic field at the maximal absorption peak position (i.e. 577 eV for Cr and 778 eV for Co) [see Fig.~3(b) and 3(c)]. While in the bulk it has been demonstrated that CoCr$_2$O$_4$ has a conical spiral spin configuration with the net magnetization contributed from three different sublattices (Co, Cr1 and Cr2), here, we find that our field-dependent XMCD data is reconciled with a new magnetic ground state described by a two-sublattice ferrimagnetic model \cite{Goodenough,Vonsovskii}. Specifically, unlike bulk, in the 2D limit the Cr1 and Cr2 sites contribute equally to the net magnetization and the spins on the remaining two magnetic sublattices of Co and Cr align anti-parallel to each other. Qualitatively, within the nearest-neighbor approximation, the Weiss molecular field on each site has contributions from both the inter-sublattice ($J_{\textrm{Co-Cr}}$) and the intra-sublattice ($J_{\textrm{Co-Co}}$ and $J_{\textrm{Cr-Cr}}$) exchange interactions. The ratio $J_{\textrm{Co-Cr}}/J_{\textrm{Cr-Cr}}$, which is a reflection of the degree of frustration, can be extracted by fitting the field-dependent XMCD data to the modified Brillouin function of this model \bibnote{For each sublattice below T$_c$, the net magnetization per site is given by the modified Brillouin function as proposed in Refs.\cite{Goodenough,Vonsovskii}: $y_i$ = $B_{S_i}(x_i)$ = $(2S_i+1)/2S_i\coth((2S_i+1)/2S_i x_i)-1/2 S_i\coth(1/2 S_i x_i)$, where  $y_i$=$M_i/M_{i0}$ and $x_i$ = $(g_i\mu_BS_i/k_B T)[H+w(-a_iM_i-M_j)], i\neq j$;  $M_i$ is the magnetization per site of the $i$th sublattice, $S_i$ is the spin magnetic moment of the $i$th sublattice, and $M_{i0}$ is the saturated value of magnetization. In the CoCr$_2$O$_4$ system $S_{\textrm{Co}}$ = $S_{\textrm{Cr}}$ = 3/2, and $g_{\textrm{(Co)}}$ = 2.2 and $g_{\textrm{(Cr)}}$ = 2.0 \cite{Altshuler}. Moreover, $w$ in $x_i$ represents the inter-sublattice Weiss constant, which is linearly proportional to the inter-sublattice exchange interaction $J_{\textrm{Co-Cr}}$;  $a_i$ is the ratio of the intra-sublattice exchange interaction with respect to the inter-sublattice value. The fit gives the value of intra- vs. inter-exchange interaction $a_{\textrm{Cr}}= J_{\textrm{Cr-Cr}}/J_{\textrm{Co-Cr}}$ = 0.49 and 0.37 for $n$ = 4 and 2 QL, respectively.}.  

\begin{table}[t]
\caption{The magnetic moment of each element (Co$^{2+}$ and Cr$^{3+}$) of all samples ($n$ = 50, 4 and 2 QL) obtained from XMCD sum rules. The net moment per CoCr$_2$O$_4$ formula unit (f.u.) is calculated as $M_{\textrm{net}} = M_{\textrm{Co}} + 2M_{\textrm{Cr}}$.}\

\centering
\begin{tabular}{c c c c}
\hline\hline
$n$ (QL) & $M_{\textrm{Co}}$ ($\mu_B$/Co$^{2+}$) & $M_{\textrm{Cr}}$ ($\mu_B$/Cr$^{3+}$) & $M_{\textrm{net}}$ ($\mu_B$/f.u.) \\ [0.6ex]
\hline
50 & 0.65 & -0.21 & 0.23 \\
4 & 0.64 & -0.15 & 0.34 \\
2 & 0.26 & -0.05 & 0.16 \\
\hline\hline
\end{tabular}
\label{Table I}
\end{table}

According to the theory proposed by Lyons, Kaplan, Dwight, and Menyuk (LKDM) \cite{LKDM_PR_1962}, in a normal spinel compound the parameter $\mu = 4J_{\textrm{BB}}S_B/3J_{\textrm{AB}}S_A$ determines spin configuration of the ground state. In particular, as shown in  Fig.~4, the ground state is a two-sublattice N\'{e}el-type collinear ferrimagnet for $\mu<0.89$ but turns into a three-sublattice spiral ferrimagnet for $0.89<\mu<1.30$; Larger values of $\mu$ indicate further enhancement of the magnetic frustration that renders the spiral ordering unstable. In our case, the Brillouin function fitting yields the experimental values of $\mu \approx$ 0.64 for 4 QL and 0.49 for 2 QL, respectively. As a result, the frustration effect becomes considerably relieved and the magnetic ground state leaves the region of spiral ferrimagnet with the propagation vector along the [110] direction \cite{Tomiyasu_PRB_2004}. In fact, in the ultrathin film geometry, the confinement along the [111] direction also breaks the translational symmetry along [110] and prevents the onset of the spiral long-range order. In addition, we can exclude the N\'{e}el-type collinear ferrimagnetic configuration as it would have required a net magnetization of $\sim$3$\mu_B$/f.u. with the overall direction following the Cr sublattice. This type of ordering is clearly in sharp variance with the observed XMCD results which show a rather small net magnetization with the direction aligned along the Co sublattice.  

To understand what kind of new magnetic ordering emerges in the ultrathin case, we recall that for an intermediate magnitude of frustration on a normal spinel lattice, Yafet and Kittel (YK) proposed another ground state which deviates from the N\'{e}el collinear configuration \cite{YK_PR_1952}. As illustrated in Fig.~4, in this model the spins on the B site are divided in two groups, each group has spins canting in an opposite way but at the same angle $\alpha_\textrm{YK}$ relative to the net magnetization direction. According to the YK theory the magnitude of the canting angle $\alpha_\textrm{YK}$ is determined by the strength of frustration. The N\'{e}el-type collinear configuration is the special case of $\alpha_\textrm{YK}$ = 0. 

In the following, we speculate on a possible mechanism for the stabilization of the YK spin configuration in (111)-oriented CoCr$_2$O$_4$ ultrathin films. First, we note that in general the YK configuration can be triggered due to structural `imperfection' of a spinel compound, i.e., the existence of tetragonal distortion \cite{Kang_JAP_2013,Chung_PRB_2013,LKDM_PRB_2_1962}, cation distribution disorder \cite{Willard_JACS_1999}, or inclusion of higher order exchange interactions \cite{Tsurkan_PRB_2003}. For our samples tetragonal distortion can be ruled out as the films are grown along the three-fold symmetry axis. Moreover, the existence of finite cation distribution disorder is excluded by our XAS results. Therefore, we suggest that the YK state is likely stabilized due to the activated additional exchange interactions. Indeed, recent LSDA+U calculations revel that \textit{J}$_{\textrm{AA}}$, which was neglected in the conventional LKDM theory, can reach $\sim$ 10\% of \textit{J}$_{\textrm{AB}}$ \cite{Ederer_PRB_2007}. Also the strength of the second nearest-neighbor B-B interaction, \textit{J}$_{\textrm{BBB}}$, is estimated $\sim$ 3\% of \textit{J}$_{\textrm{BB}}$ \cite{Dwight_JAP_1969}. Thus, by confining the films to a few QLs, the overall effect of quantum confinement is likely to enhance the relative strength of \textit{J}$_{\textrm{AA}}$ and \textit{J}$_{\textrm{BBB}}$, which shifts the magnetic energy balance towards the YK state.

\begin{figure}[t]
\includegraphics[width=0.6\textwidth]{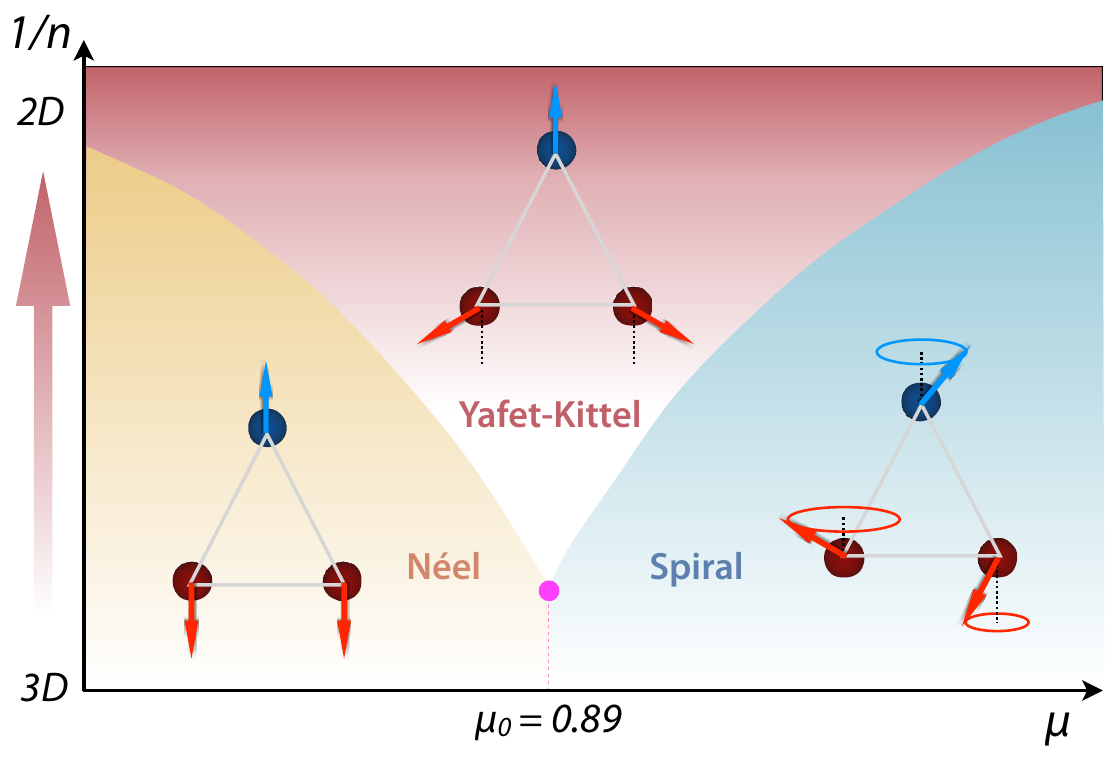}
\caption{\label{orders} The relation between inverse thickness $1/n$, parameter $\mu$ and long-range magnetic ordering in CoCr$_2$O$_4$. Note, bulk can be regarded as $n \rightarrow \infty$. The magenta solid dot stands for the possible tricritical point.}
\end{figure}  

Based on the above discussion, we propose an extended magnetic phase diagram which now includes the YK spin configuration for (111) normal spinel films as a function of $\mu$ and inverse thickness $1/n$ (dimensionality) to reflect the propensity towards 2D. As illustrated in Fig.~4, the magnetism of both bulk compounds and thick films follows the conventional LKDM theory with the magnetic ground state of either N\'{e}el or the spiral type, separated by the critical value of $\mu_0 = 0.89$. As the thickness is further reduced towards the ultrathin limit, the YK state emerges at the intermediate regime. It is interesting to point out that potentially there might exist a tri-critical point separating these three magnetic phases as marked by the magenta dot on Fig.~4. 

In summary, we report on the discovery of a new magnetic state in quasi 2D (111)-oriented CoCr$_2$O$_4$ ultrathin films. Upon the dimensionality reduction along the [111] direction, the subtle interplay among multiple exchange interactions is markedly altered and the system shifts into the intermediate region of the extended magnetic phase diagram. As a consequence of quantum confinement and the activated higher order exchange interactions, a hidden Yafet-Kittel spin configuration takes over the spiral one as the ground state. Our findings highlight the utility of dimensionally control and designed lattice topology towards novel magnetic states inaccessible in the bulk.

%%%%%%%%%%%%%%%%%%%%%%%%%%%%%%%%%%%%%%%%%%%%%%%%%%%%%%%%%%%%%%%%%%%%%
%% The "Acknowledgement" section can be given in all manuscript
%% classes.  This should be given within the "acknowledgement"
%% environment, which will make the correct section or running title.
%%%%%%%%%%%%%%%%%%%%%%%%%%%%%%%%%%%%%%%%%%%%%%%%%%%%%%%%%%%%%%%%%%%%%
\begin{acknowledgement}

The authors thank D. Khomskii, G. Fiete, X. Hu, D. D. Sarma and P. Mahadevan for numerous insightful discussions. X.L. and J.C. were supported by the Gordon and Betty Moore Foundation's EPiQS Initiative through Grant GBMF4534, and by the Department of Energy under grant DE-SC0012375. This research used resources of the Advanced Light Source, which is a Department of Energy Office of Science User Facility under Contract No. DE-AC0205CH11231. This research used resources of the Advanced Photon Source, a U.S. Department of Energy Office of Science User Facility operated by Argonne National Laboratory under Contract No. DE-AC02-06CH11357. 

\end{acknowledgement}

%%%%%%%%%%%%%%%%%%%%%%%%%%%%%%%%%%%%%%%%%%%%%%%%%%%%%%%%%%%%%%%%%%%%%
%% The same is true for Supporting Information, which should use the
%% suppinfo environment.
%%%%%%%%%%%%%%%%%%%%%%%%%%%%%%%%%%%%%%%%%%%%%%%%%%%%%%%%%%%%%%%%%%%%%
\begin{suppinfo}

Additional information regarding XAS analysis and PNR fittings. These materials are available free of charge.

\end{suppinfo}

%%%%%%%%%%%%%%%%%%%%%%%%%%%%%%%%%%%%%%%%%%%%%%%%%%%%%%%%%%%%%%%%%%%%%
%% The appropriate \bibliography command should be placed here.
%% Notice that the class file automatically sets \bibliographystyle
%% and also names the section correctly.
%%%%%%%%%%%%%%%%%%%%%%%%%%%%%%%%%%%%%%%%%%%%%%%%%%%%%%%%%%%%%%%%%%%%%
\bibliography{achemso-demo}

\begin{mcitethebibliography}{63}
\bibitem[Not()]{Note-1}
For each sublattice below T$_c$, the net magnetization per site is given by the
  modified Brillouin function as proposed in Refs.\cite{Goodenough,Vonsovskii}:
  $y_i$ = $B_{S_i}(x_i)$ = $(2S_i+1)/2S_i\coth((2S_i+1)/2S_i x_i)-1/2
  S_i\coth(1/2 S_i x_i)$, where $y_i$=$M_i/M_{i0}$ and $x_i$ =
  $(g_i\mu_BS_i/k_B T)[H+w(-a_iM_i-M_j)], i\neq j$; $M_i$ is the magnetization
  per site of the $i$th sublattice, $S_i$ is the spin magnetic moment of the
  $i$th sublattice, and $M_{i0}$ is the saturated value of magnetization. In
  the CoCr$_2$O$_4$ system $S_{\textrm{Co}}$ = $S_{\textrm{Cr}}$ = 3/2, and
  $g_{\textrm{(Co)}}$ = 2.2 and $g_{\textrm{(Cr)}}$ = 2.0 \cite{Altshuler}.
  Moreover, $w$ in $x_i$ represents the inter-sublattice Weiss constant, which
  is linearly proportional to the inter-sublattice exchange interaction
  $J_{\textrm{Co-Cr}}$; $a_i$ is the ratio of the intra-sublattice exchange
  interaction with respect to the inter-sublattice value. The fit gives the
  value of intra- vs. inter-exchange interaction $a_{\textrm{Cr}}=
  J_{\textrm{Cr-Cr}}/J_{\textrm{Co-Cr}}$ = 0.49 and 0.37 for $n$ = 4 and 2 QL,
  respectively.\relax
\mciteBstWouldAddEndPunctfalse
\mciteSetBstMidEndSepPunct{\mcitedefaultmidpunct}
{}{\mcitedefaultseppunct}\relax
\EndOfBibitem
\end{mcitethebibliography}

\end{document}